\documentclass[aps,preprint,amsmath,amssymb]{revtex4}
\usepackage{psfig}% Include figure files
\usepackage{graphicx}
\begin{document}

\title{\large \bf Are there significant scalar resonances in $B^{-}\to D^{(*)0} K^{-}
{K}^{0}$ decays? }

\author{ \bf Ron-Chou Hsieh and Chuan-Hung Chen }

%\email{physchen@mail.ncku.edu.tw}

\affiliation{ Department of Physics, National Cheng-Kung
University, Tainan 701, Taiwan }

\date{\today}% It is always \today, today,
             %  but any date may be explicitly specified

\begin{abstract}
We study the indication of large different branching ratios
between $B^{-}\to D^{(*)0} K^{-} {K}^{0}$ and $\bar{B}_{d}\to
D^{(*)+} K^{-} {K}^{0}$ observed by Belle. Interestingly, the same
situation is not found in the decays $B\to D^{(*)} K^{-}
{K}^{*0}$. If there exist no intermediate resonances in the decays
$B^{-}\to D^{(*)0} K^{-} {K}^{0}$, a puzzle will be arisen. We
find that the color-suppressed processes $B^{-}\to
D^{(*)0}a^{-}_{0}(1450)$ with $a^{-}_{0}(1450)\to K^{-} {K}^{0}$
could be one of the candidates to enhance the BRs of $B^{-}\to
D^{(*)0} K^{-} {K}^{0}$. Our conjecture can be examined by the
Dalitz plot technique and the analysis of angular dependence on
$K^{-} {K}^{0}$ state at $B$ factories.
\end{abstract}
%\pacs{13.25.Hw, 14.40.Cs}

 \maketitle
Since CP violation was discovered in $K$-meson system in 1964
\cite{CCFT}, the same thing has been also realized by Belle
\cite{Belle-CP} and Babar \cite{Babar-CP} with high accuracy in
the $B$ system. Although the new observation doesn't improve our
recognition on CP violation, it stimulates the development.
Besides the origin of CP violation, $B$ factories also provide the
chance to understand or search the uncertain states, such as the
scalar bosons $f_{0}(400-1200)$, $f_{0}(980)$, $a_{0}(980)$ and
glueball etc. Unlike pseudoscalar mesons, the scalar bosons with
wide widths are difficult to measure directly via two-body $B$
meson decays. Inevitably, the study of three body decays will
become important for extracting the quasi-two-body resonant states
from the Dalitz plot and the invariant mass distribution. By the
analyses of Dalitz plot technique, we can recognize whether there
exist unusual structures in the phase space.

Recently, Belle has observed the BRs of the decays $B\to D^{(*)}
K^{-} K^{*0}$ to be $BR(B^{-}\to D^{(*)0} K^- K^{*0})=7.5\pm
1.3\pm 1.1 \; (15.3\pm 3.1\pm 2.)\times 10^{-4}$ and
$BR(\bar{B}^{0}\to D^{(*)+} K^- K^{*0})=8.8\pm 1.1\pm 1.5 \;
(12.9\pm 2.2\pm 2.5)\times 10^{-4}$, and the decays $B\to D^{(*)}
K^{-} K^{0}$ to be $BR(\bar{B}^{-}\to D^{(*)0} K^- K^{0})=5.5\pm
1.4\pm 0.8\; (5.2\pm 2.7\pm 1.2)\times 10^{-4}$ and
$BR(\bar{B}^{0}\to D^{(*)+} K^- K^{0})=1.6\pm 0.8\pm 0.3\; (2.0\pm
1.5\pm 0.4)\times 10^{-4}$ \cite{Belle-PLB}. According to Belle's
analyses, we know that $K^{-} K^{*0}$ system has the state
$J^{P}=1^{+}$; and also, it is pointed out that the $B\to
D^{(*)}K^{-} K^{*0}$ decays in the low $K^{-} K^{*0}$ invariant
mass region can be fitted well by quasi-two-body decays $B\to
D^{(*)} a^{-}_{1}(1260)$ with $a^{-}_{1}(1260) \to K^{-} K^{*0}$.
In $K^{-}K^{0}$ system of $B^{-}\to D^{0}K^{-} K^0$ decays, by
adopting the fitting distribution $ {dN}/{d\cos\theta_{KK}}\propto
(R_L \cos^{2}\theta_{KK} + (1-R_{L}) \sin^{2}\theta_{KK})$ Belle
observed the value $R_{L}=0.97\pm 0.08$, {\it i.e.} the
$K^{-}K^{0}$ state prefers being $J^{P}=1^{-}$. Interestingly, by
looking at the data shown above, we find that the BRs of charged
$B$ decaying to $K^{-} K^{0}$ final state are much larger than
those of neutral $B$ decays. In terms of central values, the
ratios, defined by $R^{(*)}=BR(B^{-}\to D^{(*)} K^{-}
K^0)/BR(\bar{B}^{0}\to D^{(*)+} K^{-} K^0)$, could be estimated
roughly to be $3.5\ (2.6)$; however, there is no such kind of
implication on the $K^{-} K^{*0}$ final state. Therefore, if the
data display the correct behavior, there must be something
happened in $B\to D^{(*)}K^{-} K^{0}$ decays. Before discussing
the origin of the differences, we need to understand the mechanism
to produce $K^{-} K^{*0}$ and $K^{-} K^0$ systems in charged and
neutral $B$ decays.

\begin{figure}[t]
\includegraphics*[width=3.0
  in]{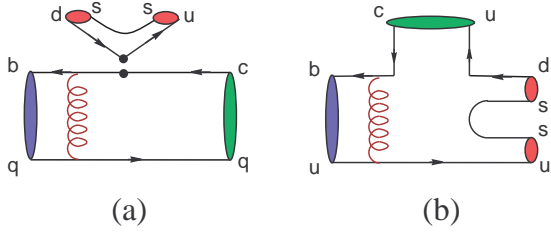} % Here is how to import EPS art
%\centerline{ \psfig{figure=brdf0.eps,height=1.6 in }
%\psfig{figure=brjf0.eps,height=1.6 in}}
\caption{Topologies for $B\to D^{(*)} K^{-} K^{(*)0}$ decays. (a)
color-allowed with $q=\text{u and d}$, and (b) color-suppressed.
}\label{three-body}
\end{figure}

Since $B\to D^{(*)} K^{-} K^{(*)0}$ decays are dictated by the
$b\to c \bar{u} d$ transition, we  describe the effective
Hamiltonian as
\begin{eqnarray}
H_{{\rm eff}}&=&\frac{G_{F}}{\sqrt{2}}V_{u}\left[ C_{1}(\mu ){\cal
O}_{1}+C_{2}(\mu ){\cal O} _{2}\right] \label{eff}
\end{eqnarray}
with ${\cal O}_{1} = \bar{d}_{\alpha} u_{\beta} \bar{c}_{\beta}
b_{\alpha}$ and  ${\cal O}_{2}^{(q)} = \bar{d}_{\alpha} u_{\alpha}
\bar{c}_{\beta} b_{\beta}$, where $\bar{q}_{\alpha}
q_{\beta}=\bar{q}_{\alpha} \gamma_{\mu} (1-\gamma_{5}) q_{\beta}$,
$\alpha(\beta)$ are the color indices, $V_{u}=V_{ud}^{*}V_{cb}$
are the products of the CKM matrix elements, $C_{1,2}(\mu )$ are
the Wilson coefficients (WCs) \cite{BBL}. As usual, the effective
WCs of $a_{2}=C_{1}+C_{2}/N_{c}$ and $a_{1}=C_{2}+C_{1}/N_{c}$,
with $N_{c}=3$ being color number, are more useful. According to
the interactions of Eq. (\ref{eff}), we classify the topologies
for $B\to D^{(*)} K^{-} K^{(*)0}$ decays to be color-allowed (CA)
and color-suppressed (CS) processes, illustrated by Fig.
\ref{three-body}, and also use the decay amplitudes
$A^{Q}(K^{-}K^{(*)0})$ to describe their decays, where $Q=C(N)$
denote the charged (neutral) $B$ decays. From the figure, we see
that both CA and CS topologies contribute to charged $B$ decays
but neutral $B$ decays are only governed by CA. We note that
because we don't consider the problem of direct CP asymmetry, the
annihilation effects, which can contribute to $\bar{B}_{d}\to
D^{(*)-} K^{-} K^{(*)0}$ decays and are usually smaller than the
CA contributions, are neglected. Although Fig. \ref{three-body}
corresponds to the picture of three-body-decay, after removing the
$s\bar{s}$ pair, the figures are related to quasi-two-body decays
if $d\bar{u}$ could form a possible bound state.

%%%%%%%%%%%%%%Figure %%%%%%%%%%%%%%%%%
%%%%%%%%%%%%%%%%%%%%%%%%%%%%%%%%%%%%%%%%%%%%%%%

Now, we have to examine that if there exists only $J^{P}=1^{-}$
state in $K^{-} K^{0}$ system, the differences between BR($B^{-}
\to D^{(*)0} K^{-} K^{0}$) and BR($\bar{B}_{d} \to D^{(*)+} K^{-}
K^{0}$) are insignificant. To be more understanding the problem,
it is useful to start from the discussion on $B\to D^{(*)} K^{-}
K^0$ processes. As mentioned before, charged $B$ decays are
governed by CA and CS while neutral $B$ decays are only from CA;
therefore, the amplitudes for $B^{-}$ and $\bar{B}^{0} $ decays
can be written as $A^{C}(K^{-}K^{0})\propto a_{1}{\cal
M}^{C}_{a}(K^{-}K^{0})+a_{2}{\cal M}^{C}_{b}(K^{-}K^{0})$ and
$A^{N}(K^{-}K^{0})\propto a_{1}{\cal M}^{N}_{a}(K^{-}K^{0})$,
where ${\cal M}^{Q}_{a(b)}$ express the hadronic transition matrix
elements for CA (CS). For simplicity, we only consider
factorizable parts. As a conjecture, if there exist significant
differences in BRs of $B^{-}$ and $\bar{B}^0$ decays, the source
should be from $a_{2}{\cal M}^{C}_{b}(K^{-}K^{0})$. In the
following, we use two examples to show that it is impossible to
only increase the BRs of $B\to D^{(*)} K^{-} K^0$ without
enhancing the BRs of $B\to D^{(*)} K^{-} K^{*0}$. Firstly, if the
decays $B\to D^{(*)} K^{-} K^0$ are only governed by
quasi-two-body decays, say $B\to D^{(*)}\rho_{X}$ with $\rho_{X}$
being arbitrary vector meson, the decay amplitudes can be
described by $A^{C}(K^{-}K^{0})\sim a_{1}f_{\rho_{X}} F^{B\to
D^{(*)} }(0)+a_{2}f_{D^{(*)}} F^{B\to \rho_{X}}(0)$ and
$A^{N}(K^{-}K^{0})\sim a_{1}f_{\rho_{X}} F^{B\to D^{(*)}}(0)$,
where $f_{D^{(*)}}$ and $f_{\rho_{X}}$ are the decay constants of
$D^{(*)}$ and $\rho_{X}$ mesons, respectively and $F^{B\to
D^{(*)}(\rho_{X})}$ denote the $B\to D^{(*)}(\rho_{X})$ form
factors. However, it is known that $|a_{2}/a_{1}|\sim 0.26$ and
$f_{\rho_{X}} F^{B\to D^{(*)}}(0) \sim f_{D^{(*)}} F^{B\to
\rho_{X}}(0)$ \cite{CY}. If the resonance is $J^{P}=1^{-}$ state,
the differences in BR between charged and neutral $B$ decays will
be slight. The same argument can be applied to $B\to D^{(*)} K^{-}
K^{*0}$ decays for $J^{P}=1^{+}$ resonant state. Hence, the
consequences are consistent with the observations of $B\to D^{(*)}
K^{-} K^{*0}$ decays. Secondly, we consider nonresonant mechanism
on $B\to D^{(*)} K^{-} K^0$. Since the situation corresponds to a
three-body phase space, we use the $K^{-}K^{0}$ invariant mass,
expressed by $\omega$, as the variable to describe the behavior of
the decay. Hence, the decay amplitudes could be written as
$A^{C}(K^{-}K^{0})\sim a_{1}F^{0\to KK}(\omega^2)F^{B\to D^{(*)}
}(\omega^2)+a_{2}f_{D^{(*)}} F^{B\to KK}(\omega^2)$ and
$A^{N}(K^{-}K^{0})\sim a_{1}F^{0\to KK}(\omega^2) F^{B\to D^{(*)}
}(\omega^2)$, and the $F^{0\to KK}$($F^{B\to KK}$) are the
associated form factors. It is known that at asymptotic region, in
terms of perturbative QCD (PQCD) the behavior of $F^{0\to KK}$ has
the power law $1/\omega^2(\ln\omega^2/\Lambda_{QCD}^2)^{-1}$
\cite{BF}. If we take the behavior of $F^{B\to D^{(*)}}$ to be
$1/(1-\omega^2/M^2_{X})^{n}$, $M_{X}$ is the pole mass and
$n=\text{1 or 2}$, we clearly see that the decay amplitudes are
suppressed at large $\omega$. Furthermore, according to the
analysis of Ref. \cite{ChenLi}, the dominant region actually is
around $\omega^2 \sim \bar{\Lambda}m_{B}$. Therefore, if $a_{2}
F^{B\to KK}(\bar{\Lambda}m_{B})$ is the source of the differences
in BR between charged and neutral $B$ decays, the similar effects
$a_{2} F^{B\to KK^*}(\bar{\Lambda}m_{B})$ will also affect the
decays $B\to D^{(*)} K^{-} K^{*0}$. Hence, unless $F^{B\to
KK^*}(\bar{\Lambda}m_{B})$ is much less than $F^{B\to
KK}(\bar{\Lambda}m_{B})$, otherwise, the CS effects should have
the similar contributions to $B\to D^{(*)} K^{-} K^{*0}$. It is
known that with the concept of two-meson wave functions \cite{MP},
the calculations of $F^{B\to KK}(\bar{\Lambda}m_{B})$ and $F^{B\to
KK^*}(\bar{\Lambda}m_{B})$ can be associated with the wave
functions of $KK$ and $KK^*$ systems. At $\omega^2\sim
\bar{\Lambda}m_{B}$ region, two-meson wave functions could be
described by two individual meson wave functions \cite{DFK}. Refer
to the derived $K$ and $K^*$ wave functions \cite{Ball}, the value
of $F^{B\to KK}(\bar{\Lambda}m_{B})/F^{B\to
KK^*}(\bar{\Lambda}m_{B})$ should be close to $F^{B\to
K}(0)/F^{B\to K^*}(0)\sim 0.8$. As a result, it is hard to imagine
that the CS effects on pure three-body decays can make large
differences in $B\to D^{(*)}K^{-} K^{0}$ processes but not in
$B\to D^{(*)}K^{-} K^{*0}$ processes.

Inspired by Belle's observation, we think the significant
differences in the BRs of $B\to D^{(*)} K^{-} K^{0}$ should not
come from $K^{-} K^0$ with $J^{P}=1^{-}$ state but with
$J^P=0^{+}$ state. This could be easily understood as follows: as
discussed before, only CS effects, $a_{2}M^{C}_{b}(K^- K^0)$,
could make a discrepancy in BRs of charged and neutral $B$ decays.
Therefore, the mechanism which dominantly contributes to CS
topology will be our candidate. If $K^- K^{0}$ has the $J^P=0^{+}$
state, due to $F^{0\to KK} \propto\; <KK,0^{+}|\bar{u}
\gamma_{\mu}d|0>$, in terms of equation of motion, we get that
$F^{0\to KK}$ is proportional to $(m_{d}- m_{u})$. That is, the
contribution of scalar state to CA topology is negligible. On the
contrary, there is no such suppression to the CS topology. Now,
the problem becomes how to produce $K^- K^0$ to be a $0^{+}$
state. By searching particle data group \cite{PDG}, we find that
the preference could be the scalar bosons $a_{0}(980)$ and
$a_{0}(1450)$ because both are isovector states and have sizable
decay rates for $a_{0}\to KK$. Hence, to realize our thought, we
propose that the decays $B^{-}\to D^{(*)0} a^{-}_{0}(980, 1450)$
with $a^{-}_{0}(980,1450)\to K^{-} K^{0}$ can satisfy the required
criterion to enhance the BRs of $B^{-}\to D^{(*)0} K^{-} K^{0}$.

Since the quark contents of scalar mesons below or near $1$ GeV
are still obscure in the literature \cite{Cheng}, for avoiding the
difficulty in estimation, we only make the explicit calculations
on $a_{0}(1450)$ which have definite composed structure of
$q\bar{q}$. If regarding $a_{0}(980)$ consists of $q\bar{q}$
state, the same estimation could be also applied \cite{Chen-PRD}.
In theory, it is known that the serious problem on two-body
nonleptonic $B$ decays comes from the calculations of hadronic
matrix elements. Since the involving processes are governed by CS
topologies, like the well known decays $B\to J/\Psi K^{(*)}$ and
$\bar{B}_{d}\to D^{(*)0} \pi^{0}$ in which nonfactorizable effects
play an important role, we adopt the PQCD approach which is
described by the convolution of hard amplitude and wave functions
\cite{LB}, can deal with the factorizable and nonfactorizable
contributions and can avoid the suffering end-point singularities
self-consistently \cite{Li}.

In the calculations of hadronic effects, the problem is how to
determine the wave functions of $D^{(*)}$ and $a_{0}$ mesons. For
$D^{(*)}$ meson wave functions, we could model them to fit with
the measured BRs of $\bar{B}_{d}\to D^{(*)0} \pi^0$ decays
\cite{Li-P}, in which $B$ meson wave function is chosen to be
coincide with the observed BRs of $B\to \pi \pi$ decays while
$\pi$ meson ones are adopted from the derivation of QCD sum rules
\cite{Ball}.  As to $a_0$ scalar meson, the spin structures of
$a_{0}$ are required to satisfy $\langle 0|\bar{u} \gamma^{\mu} d|
a^{-}_{0},p_{3}\rangle =[(m_{d}-m_{u})/m_{a_{0}}] \tilde{f}
p^{\mu}_{3}$ and $\langle 0|\bar{q} q |a_{0}\rangle = m_{a_{0}}
\tilde{f}$, where $m_{a_{0}}$ and $\tilde{f}$ are the mass and
decay constant of $a_{0}$. In order to satisfy these local current
matrix elements, the distribution amplitude for $a_{0}$ is adopted
as
\begin{eqnarray}
&& \int^{1}_{0} dx e^{ixP\cdot z} \langle 0|\bar{q}(0)_{j}
q(z)_{l}|a_{0}\rangle  \nonumber
\\ &=& \frac{1}{\sqrt{2N_{c}}}\Big\{ [\, \slash
\hspace{-0.2cm} p\, ]_{lj} \Phi_{a_{0}}(x) +m_{a_{0}}[{\bf
1}]_{lj} \Phi^{p}_{a_{0}}(x)\Big\}.\label{daf}
\end{eqnarray}
By the charge parity invariance and neglecting the effects of
light current quark mass $m_{u(d)}$, we obtain
$\Phi_{a_{0}}(x)\approx-\Phi_{a_{0}}(1-x)$ and
$\Phi^{p}_{a_{0}}(x)=\Phi^{p}_{a_{0}}(1-x)$ \cite{CZ}, and  their
normalizations are $\int^{1}_{0} dx\Phi_{a_0}(x)=0$ and
$\int^{1}_{0} dx\Phi^{p}_{a_0}(x)=\tilde{f}/2\sqrt{2N_{c}}$.
Although vector $D^{*0}$ meson carries the spin degrees of
freedom, in the $B^{-}\to D^{*0} a^{-}_{0}$ decay only
longitudinal polarization is involved. The results should be
similar to $D^{0}a^{-}_{0}$ case. Therefore, we only present the
representative formulas for $B^{-}\to D^{0} a^{-}_{0}$ decay.
Hence, the decay amplitude for $B^{-}\to D^{0} a^{-}_{0}$ is read
as
\begin{eqnarray*}
A &=&  V_{u} \Big[f_{D }{\cal F}^{e}_{D^{0} a_{0}} +{\cal
M}^{e}_{D^{0} a_{0}} \Big]
\end{eqnarray*}
where ${\cal F}^{e}({\cal M}^{e})$ is the factorized
(non-factorized) emission hard amplitudes. According to Eq.
(\ref{daf}), the typical hard functions are expressed as
\begin{eqnarray}
 {\cal F}^{e}_{D^{0} a_{0}}&=&\eta \int_{0}^{1} dx_{1} dx_{3}
\int_{0}^{\infty} b_{1}db_{1}b_{3}db_{3}\Phi _{B}( x_{1},b_{1})
\nonumber\\
&&  \Big\{ \Big[(\zeta_{1} + \zeta_{2} x_{3})\Phi_{a_{0}}(x_{3})
+r_{a}(1-2x_{3})\Phi^{p}_{a_{0}}(x_{3})\Big] \nonumber \\
&& {\cal E}_{e}(t^{1}_{e})+r_{a}\Big(2 \Phi^{p}_{a_{0}}(x_3)-r_{a}
\Phi_{a_{0}}(x_{3})\Big)\nonumber \\ && \times {\cal
E}_{e}(t^{2}_{e}) \Big\},
\label{fe}\\
%%%%%%%%%%%%%%%%%%%%%%%%%%%%%%%%
{\cal M}^{e}_{D^{0} a_{0}}&=&2\eta \int_{0}^{1} d[x]
\int_{0}^{\infty} b_{1}db_{1} b_{3}db_{3} \Phi _{B}(
x_{1},b_{1})\Phi_{D}(x_{2})
\nonumber\\
&& \Big\{ \Big[-\Big((1-r^2_{a})x_{2}+(1-r^2_{D})x_{3}\Big)
\Phi_{a_{0}}(x_{3})\nonumber \\
&& +r_{a}x_3\Phi^{p}_{a_{0}}(x_{3})\Big] {\cal
E}^{1}_{d}(t^{1}_{d})+\Big[\Big((\zeta_{1}-r^{2}_{a})(1-x_{2})\nonumber
\\ &&+r^{2}_{a}x_{3}
-r^{2}_{D}x_{2}\Big)\Phi_{a_{0}}(x_{3})-r_{a}x_{3}\Phi^{p}_{a_{0}}(x_3)\Big]\nonumber
\\&& \times {\cal E}^{2}_{d}(t^{2}_{d}) \Big\}, \label{me}
\end{eqnarray}
%%%%%%%%%%%%%%%%%%%%%%%%%%%%%%%%%%
with $\eta=8\pi C_{F} M^{2}_{B}$, $r_{a(D)}=m_{a_{0}(D)}/M_{B}$,
$\zeta_{1}=1-r^2_{a}-r^{2}_{D}$, $\zeta_{2}=\zeta_{1}-r^{2}_{D}$,
${\cal E}^{i}_{e}(t^{i}_{e})=$ $\alpha_{s}(t^{i}_{e})$
$a_{2}(t^{i}_{e})$ $Su_{B+a_{0}}(t^{i}_{e})$ $h_{e}(\{x\},\{b\})$
and ${\cal
E}^{i}_{d}(t^{i}_{d})=\alpha_{s}(t^{i}_{d})(C_{2}(t^{i}_{d})/N_{c})$
$Su(t^{i}_{d})_{B+D+a_{0}}$ $h_{d}(\{x\},\{b\})$. $t^{1,2}_{e,d}$,
$Su$ and $h_{e,d}$ denote the hard scales of $B$ decays, Sudakov
factors and hard functions which are arisen from the propagators
of gluon and internal valence quark, respectively. Their explicit
expressions can be found in Ref. \cite{Chen-Li}.

For numerical estimations, the $B$, $D^{(*)0}$ and $a_{0}$ meson
wave functions are simply chosen as
\begin{eqnarray}
\Phi_{B}(x,b)&=& N_{B}x^{2}(1-x)^{2} \exp\Big[-\frac{1}{2}\Big(
\frac{x
\,m_{B}}{\omega_{B}}\Big)^{2}-\frac{\omega_{B}^{2}b^{2}}{2}
\Big]\nonumber \\
 \Phi_{D^{(*)}}(x)&= & {3 \over \sqrt{2N_{c}}}
f_{D^{(*)}}x(1-x)[1+0.7 (1-2x)], \nonumber \\
 \Phi_{a_{0}}(x)
&=& {\tilde{f} \over 2\sqrt{2N_{c}}}
\left[6x(1-x)C^{3/2}_{1}(1-2x)\right],\nonumber \\
 \Phi^{p}_{a_{0}}(x)&=& \frac{\tilde{f}}{2\sqrt{2N_{c}}} %\bigg\{
\Big[3(1-2x)^{2}\Big],
%+ G_{1}^{p} (1-2x)^{2} \nonumber
%\\
%&& \times \left[C^{3/2}_{2}(1-2x)-3 \right]+ G^{p}_{2}
%C^{1/2}_4(1-2x)
%\bigg\},
 \label{wavf}
\end{eqnarray}
where $N_{B}$ is determined by the normalization of $\int^{1}_{0}
dx \Phi_{B}(x,0)=f_{B}/(2\sqrt{2N_{c}})$ and $C^{3/2}_{1}(y)$ is
the Gegenbauer polynomial. The values of theoretical inputs are
set as: $\omega_{B}=0.4$, $f_{B}=0.19$, $f_{D^{(*)}}=0.20\,
(0.22)$, $\tilde{f}=0.20$, $m_{B}=5.28$ and $m_{D^{(*)}}=1.87\,
(2.01)$ GeV. By the taken values and using Eq. (\ref{fe}) with
excluding WC of $a_{2}$, we immediately obtain the $B\to
a_{0}(1450)$ form factor at large recoil to be $0.44$. The result
is quite close to the value $0.46$ which is estimated by the
light-cone sum rules \cite{Chernyak}. Hence, the magnitudes of the
considered hard functions are given
in Table \ref{tablehf}. %%%%%%%%%%%%%%
%%%Table
%%%%%%%%%%%%%%%%%%%%%%%%
\begin{table}[htb]
\caption{Hard functions (in units of $10^{-2}$) for $B^{-}\to
D^{(*)0} a^{-}_{0}(1450)$ decays with $\omega_{B}=0.4$,
$f_{B}=0.19$, $f_{D^{(*)}}=0.2\; (0.22)$ and $\tilde{f}=0.20$ GeV.
}\label{tablehf}
\begin{ruledtabular}
\begin{tabular}{cccc}
  $ \ \ f_{D}{\cal F}^{e}_{D^{0} a_{0}} \  \ $ & $\ \ {\cal
M}^{e}_{D^{0} a_{0}}\ \ $ & $\ \ f_{D^*}{\cal
F}^{e}_{D^{*0} a_{0}} \ \ $ & $\ \ {\cal M}^{e}_{D^{*0} a_{0}}\ \ $  \\
\hline   $-1.42$  & $-2.22 + i 0.97$  & $-1.56$ & $-2.39+i 1.06$
\end{tabular}
\end{ruledtabular}
\end{table}
%%%%%%%%%%%%%%%%%%%%%%%%%%%%%%%%%%%%%%%
Consequently, the BRs of $B^{-}\to D^{(*)0} a^{-}(1450)$ are
obtained to be $8.21 \;(9.34) \times 10^{-4}$. Since the
predictions of PQCD on the BRs of $\bar{B}\to D^{0} \pi^0$
\cite{Li-P} and $B\to J/\Psi K^{(*)}$ and the helicity amplitudes
of $B\to J/\Psi K^*$ \cite{Chen-Lett} are consistent with the
current experimental data, with the same approach, our results
should be reliable. Furthermore, the BR products of $Br(B^{-}\to
D^{(*)0}a^{-}_{0}(1450))\times Br(a^{-}_{0}(1450)\to K^-
K^0)\approx 1.81\; (2.05)\times 10^{-4}$ with
$Br(a^{-}_{0}(1450)\to K^- K^0)\sim 0.22$ \cite{PDG}. Clearly, the
contributions of quasi-two-body decays to $B^{-}\to D^{(*)0} K^{-}
K^0$ modes are close to the pure three-body decays $\bar{B}\to
D^{(*)+} K^- K^0$ \cite{Hou}.

In summary, we have investigated that when a proper scalar meson
is considered, the BRs of $B^{-}\to D^{(*)0} K^{-} {K}^{0}$ will
deviate from those of $\bar{B}_{d}\to D^{(*)+} K^{-} {K}^{0}$.
Although we only concentrate on $a_{0}(1450)$, the same discussion
is also applicable to $a_{0}(980)$. Since our purpose is just to
display the importance of scalar boson on the decays $B^{-}\to
D^{(*)0} K^{-} {K}^{0}$, we don't consider the theoretical
uncertainties at this stage. And also, we neglect discussing the
interfering effects of resonance and non-resonance. It is worthful
to mention that by using powerful Dalitz plot technique, many
scalar mesons in charm decays have been observed by the
experiments at CLEO \cite{CLEO}, E791 \cite{E791}, FOCUS
\cite{FOCUS}, and Babar \cite{Babar}. Expectably, the significant
evidences of scalar productions should be also observed at $B$
factories. In addition, since the scalar meson doesn't carry spin
degrees of freedom, there is no specific direction for
$K^{-}K^{0}$ production so that we should see a different angular
distribution of $K^{-}K^{0}$, such as the coefficient associated
with the term $\sin^{2}\theta_{KK}$ in the fitting distribution
mentioned early will be enhanced. Hence, with more data
accumulated, our conjecture can be examined by the Dalitz plot and
the analysis of
angular dependence on $K^{-} {K}^{0}$ state.\\

\noindent {\bf Acknowledgments:}

The author would like to thank C.Q. Geng, H.N. Li and  H.Y. Cheng
for their useful discussions. This work was supported in part by
the National Science Council of the Republic of China under Grant
No. NSC-92-2112-M-006-026 and by the National
Center for Theoretical Sciences of R.O.C.. \\

\end{document}